\documentclass[conference,a4paper]{IEEEtran}
\usepackage{amssymb}
\usepackage{amsmath}
\usepackage{amsthm}
\usepackage{subfigure, epsfig}
\usepackage{pstricks}
\usepackage{pstricks-add}
\usepackage{pst-plot}
\usepackage{pseudocode}
\usepackage[ruled,boxed,commentsnumbered]{algorithm2e}
\usepackage{graphicx}
\usepackage{xcolor}
\newtheorem{thm}{Theorem}
\newtheorem{lem}{Lemma}

\newtheorem{theorem}{Theorem}
\newtheorem{definition}[theorem]{Definition}

\newcommand{\ul}{\underline}
\newcommand{\ol}{\overline}

\newcommand{\mc}{\mathcal}

\newcommand{\ds}{\displaystyle}

\begin{document}

\sloppy

\title{Coded Power Control: Performance Analysis}

 \author{
   \IEEEauthorblockN{Benjamin Larrousse and Samson Lasaulce}
   \IEEEauthorblockA{L2S (CNRS -- Sup\'{e}lec -- Univ. Paris Sud 11),~91191 Gif-sur-Yvette, France\\
     Email: \{larrousse, lasaulce\}@lss.supelec.fr}
 }
\maketitle

\begin{abstract} In this paper, we introduce the
general concept of coded power control (CPC) in a particular setting of the interference channel. Roughly, the idea of CPC consists in embedding information (about a random state) into the transmit power levels themselves: in this new framework, provided the power levels of a given transmitter can be observed (through a noisy channels) by other transmitters, a sequence of power levels of the former can therefore be used to coordinate the latter. To assess the limiting performance of CPC (and therefore the potential performance brought by this new approach), we derive, as a first step towards many extensions of the present work, a general result which not only concerns power control (PC) but also any scenario involving two decision-makers (DMs) which communicate through their actions and have the following information and decision structures. We assume that the DMs want to maximize the average of an arbitrarily chosen instantaneous payoff function which depends on the DMs' actions and the state realization. DM 1 is assumed to know non-causally the state (e.g., the channel state) which affects the common payoff while the other, say DM 2, has only a strictly causal knowledge of it. DM 1 can only use its own actions (e.g., power levels) to inform DM 2 about its best action in terms of payoff. Importantly, DM 2 can only monitor the actions of DM 1 imperfectly and DM 1 does not observe DM 2. The latter assumption leads us to exploiting Shannon-theoretic tools in order to generalize an existing theorem which provides the information constraint under which the payoff is maximized. The derived result is then exploited to fully characterize the performance of good CPC policies for a given instance of the interference channel.
\end{abstract}

\section{Introduction}
\label{sec:introduction}

Consider two decision-makers (DMs) and that each of them has to select actions or take decisions repeatedly to reach a certain objective say to maximize an average payoff function. Furthermore, assume that there might be an interest for them in exchanging information e.g., about a random system state which can affect their payoff but that no dedicated communication channel is available for this purpose. Therefore, the only way to communicate for a DM is to use his own actions. Although the idea of communicating through actions seems to be quite natural and is in fact used more or less implicitly in real life scenarios e.g., in economics (see \cite{sims-jme-2003}), it appears that, apart from a few exceptions focused on specific problems of control (see e.g., \cite{grover-phd-2011}), it has obviously not penetrated yet engineering problems and definitely not wireless communications. It turns out that important wireless problems such as power control (PC) or radio resource allocation can draw much benefits from being revisited from the new perspective of communication through actions. Because of its importance and ability to easily illustrate the proposed approach, the problem of PC in interference networks has been selected for the application of the main and general result derived in this paper; note that the latter concerns any decision-making problem which has the same structure (see Sec. \ref{sec:problem}) and generalizes \cite{Gossner-2006} and \cite{cuff-itw-2011}.

In the context of PC, the DMs are transmitters (Txs) and the system state is typically given by the state of the communication channel between the Txs and receivers (Rxs); we will use the term DM (resp. Tx) when the general case (resp. the specific case of PC) is concerned. Quite often, each Tx possesses a partial knowledge of the channel state and, in general, there is an incentive for the Txs to exchange the corresponding knowledge between them. Coded PC assumes that this knowledge is transferred from one Tx to another (or others) by encoding the information of the former into a sequence of power levels which are observed by the latter. In this paper, we assume two DMs, that they have a common payoff, and that there is one DM (DM 1) which is informed with the current realization of the state and possibly those associated with the coming stages. DM 2 is only informed of the state in a strictly causal manner and can observe his on power levels. The considered scenario is, in particular, relevant in cognitive radio (CR) settings. In typical CR scenarios, the primary Tx is assumed to be passive and the secondary Tx adapts to what it observes. But, it might be of interest to design primary Txs which coordinate in an active manner the usage of radio resources, which is exactly what coded power control (CPC) allows; one of the salient features of CPC is that interference can be managed directly in the radio-frequency domain and does not require baseband detection or decoding, which is very useful in heterogeneous networks. Another body of works which can be mentioned is given by works on distributed PC and especially those on best response dynamics (BRD) algorithms which include the original iterative water-filling algorithm \cite{yu-jsac-2002}. Existing BRD algorithms implementations for PC (see e.g., \cite{lasaulce-book-2011}\cite{zappone-comlett-2011}\cite{bacci-tsp-2013}) typically assume SINR (signal-to-noise plus interference ratio) feedback and individual channel state information (CSI) and do not exploit the key idea of communicating through the power levels. Encoding power levels allows one to construct PC policies possessing at least three salient features which are generally not available for BRD-based PC: there is no convergence problem and this whatever the payoff functions; efficient solutions (e.g., in terms of sum-payoff) can be obtained; both the cases of discrete and continuous power levels can be easily treated. Since we focus on optimal PC policies and make the choice of an asymmetric information structure whereas BRD algorithms rely on a symmetric one, no explicit comparison with BRD algorithms is conducted but CPC can be applied to symmetric scenarios as well.

\section{Problem statement}
\label{sec:problem}

Consider two DMs which want to coordinate through their actions. Let $\mathcal{X}_j$, $|\mathcal{X}_j| < \infty$, the action alphabet of DM $j \in \{1,2\}$, and $\mathcal{X}_0$, $|\mathcal{X}_0|<\infty$, the random state alphabet. The states are assumed to be i.i.d. and generated from a random variable $X_0$ whose realizations are in $\mc{X}_0$ and distribution is denoted by $\rho$. Note that the finiteness assumption is not only realistic (e.g., power levels are discrete in modern cellular systems) but also allows the continuous case to be treated by using classical arguments \cite{Cover:2006:EIT:1146355}. The strategies of DM $1$ and $2$ are sequences of mappings, $(\sigma_i, \tau_i)_{i\geq1}$,  which are respectively defined by:
\begin{equation}\label{eq:strategies}
\left\{
\begin{array}{ccccc}
\sigma_i & : & \mathcal{X}_0^T \times \mathcal{X}_1^{i-1} & \rightarrow & \mathcal{X}_1\\
\tau_i & : & \mathcal{X}_0^{i-1} \times \mathcal{Y}^{i-1} \times \mathcal{X}_2^{i-1} & \rightarrow & \mathcal{X}_2
\end{array}
 \right.
\end{equation}
where $T$ is the total number of stages, $i\in \{1,...,T\}$, and $\mc{Y}$, $|\mc{Y}| < \infty$, is the observation alphabet of DM $2$. The definition of the strategy for DM 1 indicates that we assume a non-causal knowledge of the state. The most typical situation in PC is to assume that two phases are available (training phase, action phase) and one state is known in advance to adjust the power level. This special case can be obtained by setting $T=2$ that is, $i\in\{1,2\}$. There are many reasons why we consider here that $T$ might be greater than two. We will only provide three of them, which better explains how the non-causality assumption should be understood. First, the result derived in Sec. \ref{sec:main-analytical-result} can be used for a large variety of settings and not only PC. Second, the proposed approach can be applied to the case where the state is not i.i.d. (e.g., to the $B-$stage block i.i.d. case, $B\geq1$). Indeed, there exist wireless communication standards which assume the channel to be constant over several time-slots and the proposed approach suggests that gains can be obtained by varying the power level from time-slot to time-slot even if the channel is constant. Third, it becomes more and more common to exploit the forecasted trajectory of a mobile user to optimize the system \cite{fourestie-patent-2007}, which makes our approach relevant when the channel state is interpreted as the path loss. Concerning the chosen definition for the strategy of DM 2, several comments are in order. First, note that DM 2 is not assumed to monitor actions of DM 1 perfectly. Rather, they are monitored through an observation channel which is assumed to be discrete, memoryless, and to verify $P(y|x_0,x_1,x_2) = \Gamma(y|x_1)$, where $y \in \mc{Y}$ is a realization of the channel output associated with the input $(x_0,x_1,x_2)$. Second, note that, the strategy of DM 2 is defined such that it can choose an action at every stage and not only at the end of a block or sequence of stages as it would be for a classical block decoder. Therefore, contrarily to \cite{Gossner-2006}, DM 1 does not need to observe the actions of DM 2 and DM 2 has only access to imperfect observations of the actions chosen by DM 1. Interestingly, we will see that the fact that DM 1 does not observe DM 2 induces no performance loss in terms of payoff.

The instantaneous or stage payoff function for the DMs is denoted by $w(x_0,x_1,x_2)$. Since the state is random, we will consider as general case the problem of reaching a certain performance level in terms of expected payoff $\mathbb{E}[w] = \sum_{(x_0,x_1,x_2)} P(x_0,x_1,x_2) w(x_0,x_1,x_2)$. Roughly, the task of DM 1 is to maximize the expected payoff by finding the best tradeoff between reaching a good payoff for the current stage and revealing enough information about the future realizations of the state to coordinate for the next stages. The ability for two DMs to coordinate their actions i.e., to reach a certain value for the expected payoff can be translated in terms of joint distribution over $\mc{X}_0\times\mc{X}_1\times\mc{X}_2$, which leads us to the notion of implementable distribution \cite{Gossner-2006}.

\begin{definition}[Implementability] The distribution $\ol{Q}(x_0,x_1,x_2)$ is implementable if there exists a pair of strategies $(\sigma_i, \tau_i)_{i\geq1}$ such that as $t \rightarrow +\infty$ we have for all $(x_0,x_1,x_2)$,
\begin{equation}
\frac{1}{t} \sum_{i=1}^{t} \sum_{y}  P_{X_{0,i}, X_{1,i}, X_{2,i}, Y_i}(x_0,x_1,x_2,y) \rightarrow \ol{Q}(x_0,x_1,x_2)
\end{equation}
where $P_{X_{0,i},X_{1,i},X_{2,i},Y_i}$ is the joint distribution induced by $(\sigma_i, \tau_i)_{i\geq1}$ at stage $i$.
\end{definition}

Importantly, note that, since the expectation operator is linear, a certain value for $\mathbb{E}[w]$ is reachable if and only if there exists an implementable distribution. The goal of the next section is precisely to characterize the set of reachable expected payoffs $\mathbb{E}_Q[w] = \sum_{(x_0,x_2,x_2,y)}  \ol{Q}(x_0,x_1,x_2) \Gamma(y|x_1) w(x_0,x_1,x_2)$, which thus amounts to characterizing the set of implementable distributions over $\mc{X}_0 \times \mc{X}_1 \times \mc{X}_2$.

\section{Main analytical result}
\label{sec:main-analytical-result}

Notation: $\Delta(\mc{A})$ will stand for the set of
distributions over the generic discrete set $\mc{A}$. Using this notation, the main analytical result of this paper can be stated.

\begin{thm}\label{theo:info-constraint} Let $\ol{Q} \in \Delta(\mathcal{X}_0 \times \mathcal{X}_1 \times \mathcal{X}_2)$ with $\sum_{(x_1,x_2)}\ol{Q}(x_0,x_1,x_2) = \rho(x_0)$. The distribution $\ol{Q}$ is implementable if and only if there exists  $Q \in \Delta(\mathcal{X}_0 \times \mathcal{X}_1 \times \mathcal{X}_2 \times \mathcal{Y})$ which verifies the following information constraint:
\begin{equation}
\label{eq:information-constraint}
I_{Q} (X_0 ; X_2) \leq I_{Q} (X_1;Y |X_0,X_2)
\end{equation}
where the arguments of the mutual information $I_{Q}(.)$ are defined from $Q$ and $Q(x_0,x_1,x_2,y) = \ol{Q}(x_0,x_1,x_2) \Gamma(y|x_1)$.
\end{thm}

\subsection{Proof of Theorem 1}
\label{sec:proof-main-result}

\begin{IEEEproof}[Converse proof]  We first start with providing a lemma which is used at the end of the proof and concludes the section.

\begin{lem}\label{lemma:convexity-of-phi} The function $\Phi : Q  \mapsto I_{Q} (X_0 ; X_2) - I_{Q} (X_1;Y |X_0,X_2)$ is convex over the set of distributions $Q \in \Delta(\mc{X}_0 \times \mc{X}_1 \times \mc{X}_2 \times \mc{Y})$ that verify $\sum_{(x_1,x_2,y) Q(x_0,x_1,x_2,y)} = \rho(x_0)$ and  $Q(x_0,x_1,x_2,y) = \Gamma(y|x_1) P(x_0,x_1,x_2)$, with $\rho$ and $\Gamma$ fixed.
\end{lem}

\setlength{\belowdisplayskip}{0pt} \setlength{\belowdisplayshortskip}{0pt}
\setlength{\abovedisplayskip}{0pt} \setlength{\abovedisplayshortskip}{0pt}

\begin{IEEEproof}[Proof of Lemma \ref{lemma:convexity-of-phi}] The function $\Phi$ can be rewritten as $\Phi(Q) = H_Q(X_0) - H_Q(Y,X_0 | X_2) + H_Q(Y|X_0,X_2,X_1)$. The first term $H_Q(X_0) = -\sum_{x_0} \rho(x_0) \log \rho(x_0)$ is a constant w.r.t. $Q$. The third term is linear w.r.t. $Q$ since,  with $\Gamma$ fixed,

\begin{multline}
 H_Q(Y|X_0,X_2,X_1) = \\
  - \sum_{x_0,x_1,x_2,y} Q(x_0,x_1,x_2,y) \log P(y|x_0,x_1,x_2) \\
  = - \sum_{x_0,x_1,x_2,y} Q(x_0,x_1,x_2,y) \log \Gamma(y|x_1)
\end{multline}
It is therefore sufficient to prove that $H_Q(Y,X_0 | X_2)$ is concave. Let $\lambda_1 \in [0,1]$, $\lambda_2 = 1 - \lambda_1$, $(Q_1,Q_2) \in  \Delta^2(\mc{X}_0 \times \mc{X}_1 \times \mc{X}_2 \times \mc{Y})$ and $Q=\lambda_1 Q_1 + \lambda_2 Q_2$. By using the standard notation $A^0 = \emptyset$, $A^n = (A_1, ..., A_n)$, we have that:
\begin{align}
   & H_Q(Y,X_0 | X_2) = - \sum_{x_0,x_2,y} \bigg( \sum_{x_1,i} \lambda_i Q_i(x_0,x_1,x_2,y) \bigg) \nonumber \\
  &  \log \left[ \frac{\sum_{x_1,i} \lambda_i Q_i(x_0,x_1,x_2,y)}{ \sum_{i} \lambda_i P_{X_2}^{Q_i}(x_2)} \right] \\
  &  =  - \sum_{x_0,x_2,y} \bigg(\sum_i \lambda_i  \sum_{x_1} Q_i(x_0,x_1,x_2,y) \bigg) \nonumber \\
   &  \log \left[ \frac{\sum_i \lambda_i  \sum_{x_1} Q_i(x_0,x_1,x_2,y)}{ \sum_{i} \lambda_i P_{X_2}^{Q_i}(x_2)} \right] \\
  & \geq - \sum_i \lambda_i \sum_{x_0,x_2,y} \bigg(  \sum_{x_1} Q_i(x_0,x_1,x_2,y) \bigg) \nonumber \\
   & \log \left[ \frac{\lambda_i  \sum_{x_1} Q_i(x_0,x_1,x_2,y)}{\lambda_i P_{X_2}^{Q_i}(x_2)} \right] \\
& = - \sum_i \lambda_i \sum_{x_0,x_2,y} \bigg(  \sum_{x_1} Q_i(x_0,x_1,x_2,y) \bigg) \nonumber \\
& \log \left[ \frac{\sum_{x_1} Q_i(x_0,x_1,x_2,y)}{ P_{X_2}^{Q_i}(x_2)} \right] \\
& =  \lambda_1 H_{Q_1}(Y,X_0 | X_2) + \lambda_2 H_{Q_2}(Y,X_0 | X_2)
\end{align}
where the inequality comes from the log sum inequality \cite{Cover:2006:EIT:1146355}.
\end{IEEEproof}

Now we want to prove that if $\ol{Q}$ is implementable, then $Q$ has to verify the information constraint. Assuming $\ol{Q}$ is implementable means that there exists $(\sigma_i, \tau_i)_{i\geq1}$ such that the empirical distribution $P^{(t)}_{X_0,X_1,X_2,Y}(.)  =  \frac{1}{t} \sum_{i=1}^{t} P_{X_{0,i}, X_{1,i}, X_{2,i}, Y_i}(.)$ can be made arbitrarily close to $Q$; this argument is used at the end of the proof. We have:

\begin{align}
& \sum_{i=1}^{t} I_{P_{X_{0,i},X_{1,i},X_{2,i},Y_i}} ( X_0 ; X_{2}) = \sum_{i=1}^{t} I ( X_{0,i} ; X_{2,i}) \\
\stackrel{(a)}{=} & H(X_0^t) - \sum_{i=1}^{t} H(X_{0,i}| X_{2,i})  \\
 = &  H(X_0^t,Y^t,X_2^t) - H(Y^t,X_2^t | X_0^t) - \sum_{i=1}^{t} H(X_{0,i}| X_{2,i}) \\
\leq &  H(X_0^t,Y^t,X_2^t) - H(Y^t | X_0^t) - \sum_{i=1}^{t} H(X_{0,i}| X_{2,i})
\end{align}
\begin{align}
\leq &  H(X_0^t,Y^t,X_2^t) - H(Y^t | X_0^t,X_{1}^t,X_{2}^t) - \sum_{i=1}^{t} H(X_{0,i}| X_{2,i}) \\
\stackrel{(b)}{=} &   H(X_0^t,Y^t,X_2^t) - \sum_{i=1}^{t} H(X_{0,i}| X_{2,i}) \nonumber \\
& - \sum_{i=1}^{t} H(Y_{i} | X_{0,i},X_{1,i},X_{2,i}) \\
\stackrel{(c)}{\leq} &  \sum_{i=1}^{t} H(X_{0,i},Y_{i},X_{2,i} | X_{2,i}) - H(X_{0,i}| X_{2,i}) \nonumber \\
&- H(Y_{i} | X_{1,i},X_{0,i},X_{2,i})  \\
= & \sum_{i=1}^{t} H(X_{0,i},Y_{i} | X_{2,i}) - H(X_{0,i}| X_{2,i}) \nonumber  \\
&- H(Y_{i} | X_{1,i},X_{0,i},X_{2,i}) \\
= & \sum_{i=1}^{t} I(X_{1,i};Y_{i} | X_{0,i},X_{2,i}) \\
= &  \sum_{i=1}^{t} I_{P_{X_{0,i},X_{1,i},X_{2,i},Y_i}} (X_{1} ; Y | X_0,X_2)
\end{align}

where: (a) comes from the fact that $(X_{0,i})_i$ is i.i.d. and the chain rule for entropy; (b) holds because the observation channel from DM 1 to DM 2 is assumed to be discrete and memoryless namely, $P(y^t|x_0^t,x_1^t,x_2^t)=\prod_{i=1}^{t} p(y_{i}|x_{0,i},x_{1,i},x_{2,i})$; (c) holds by the chain rule and because $X_{2,i}$ is a deterministic function of the past: $X_{2,i}= \tau_i \left(X_{0,1},Y_{1},X_{2,1},\dots,X_{0,i-1},Y_{i-1},X_{2,i-1}\right)$. Now, since $\Phi$ is convex (by Lemma \ref{lemma:convexity-of-phi}), we know that
\begin{multline} I_{P^{(t)}_{X_0,X_1,X_2,Y}} (X_{1} ; Y | X_0, X_2)
- I_{P^{(t)}_{X_0,X_1,X_2,Y}} ( X_0 ; X_{2}) \geq  \\
\frac{1}{t} \sum_{i=1}^{t}  I_{P_{X_{0,i},X_{1,i},X_{2,i},Y_i}} (X_{1} ; Y | X_0, X_2) \\
 - I_{P_{X_{0,i},X_{1,i},X_{2,i},Y_i}} ( X_0 ; X_{2})
\end{multline}

The converse follows by observing that the first term of the above inequality can be made arbitrarily close to $I_{Q} (X_1;Y |X_0,X_2) - I_{Q} (X_0 ; X_2)$ and the second term has been proven to be non-negative. \hfill 
\end{IEEEproof}

\begin{IEEEproof}[Implementability (sketch)]
 The goal here is to prove that if the information constraint is verified for $Q^*$, then an implementable pair of strategies $(\sigma_i, \tau_i)_{i\geq1}$ can be found. Therefore, in contrast with the converse, finding a particular code such as a block code with long codewords is sufficient, which allows one to reuse the standard machinery for the transmission of distorted sources. Assume $T = n B$ large where $n$ is the codeword length and $B$ the number of blocks. Denote $b$ as the block index. The methodology is the following: construct a source codebook and a channel codebook by choosing each symbols of each sequences in the codebooks independently using the same distribution $Q^*$ (more precisely marginal distributions of $Q^*$). DM 1 then uses joint typicality to find in the source codebook the sequence of actions of DM 2 for block $b+1$, and sends the corresponding channel codeword (the one with the same index). DM 2 receives a sequence $y^n[b]$ through the observation channel and decodes the index chosen by DM 1 via joint typicality of the four sequences on block $b$. He uses this index to find in the source codebook his sequence of actions for block $b+1$. At last, for block $b=0$, DM 2 chooses an arbitrary codeword which is known to DM 1. There is an error if we don't have existence and/or unicity of these codewords. The probability of error is made arbitrarily small thanks to the information constraint \eqref{eq:information-constraint} (the analysis, although not trivial, is standard and need the well known Markov Lemma and Packing Lemma). Under this setting, as $Q^*$ meets the information constraint, the empirical distribution $Q^{(T)}$ which is induced by this separate source channel coding i.e.,
\begin{multline} Q^{(T)}(v)=\frac{1}{n B} \big[ \mathcal{N}\big(v \; | \; x_0^n(0),x_1^{n}(0),x_2^n(0),y^{n}(0) \big) \\
+ \sum_{b=1}^{B-1} \mathcal{N}\big(v \; | \; x_0^n(b),x_1^{n}(b),x_2^n(b),y^{n}(b) \big) \big] 
\end{multline}
converges to $Q^*$, where $\mc{N}(v|v^n)$ is a notation for counting the occurrences of $v$ in $v^n$, $v = (x_0,x_1,x_2,y)$ here. The proof of this involves definitions of typical sets and the triangle inequality.
\end{IEEEproof}

\subsection{Comments on Theorem 1}
\label{sec:comments-main-result}
Theorem 1 can be interpreted as follows. DM 2's actions (represented by $X_2$) correspond to a joint source-channel decoding operation with distortion on the information source (which is represented by $X_0$). To be reachable, the distortion rate has to be less than the transmission rate allowed by the channel whose input and output are respectively represented by $X_1$ and $Y$. Therefore, the pair $S=(X_0,X_2)$ seems to play the same role as the side information in channels with state. Indeed, the implementability proof shows that DM 1 uses in particular $(x_0^n(b), \widehat{x}_2^n(b))$ while DM 2 uses $(x_0^n(b), x_2^n(b))$.  Asymptotically, the encoder (DM 1) and decoder (DM 2) have the same side information (which explains by the way the fact DM 1 does not need to observe DM 2 does not induce any performance loss). Furthermore, note that $x_0^n(b+1)$, which plays the role of the message to be encoded, is independent of the side information. Classical coding schemes (such as block Markov coding) can thus be re-exploited. However, the above arguments fails for the converse proof which has to deal with arbitrary coding schemes or strategies. It can no longer be assumed that the side information be independent of the information source vector. This is one of the reasons why the converse proof has to be rethought. Another reason is that classical results (such as Fano's inequality) which rely on block decoding are not exploitable anymore since DM 2 has to be able to act (to decode) at any stage or time instance.

As another type of comments on Theorem \ref{theo:info-constraint}, it can be noted that the information constraint has a very attractive property: the problem of maximizing the expected payoff
takes a particularly simple form. Indeed, by defining a one-to-one mapping between the quadruplets $(x_0,x_1,x_2,y)$ and the finite set $\{1,2,...,L\}$, $L = |\mc{X}_0 \times \mc{X}_1 \times \mc{X}_2 \times \mc{Y} |$, the optimization problem of interest can be described as follows:
\begin{equation}
\begin{array}{crcl}
\text{minimize} & -\mathbb{E}_{\ul{q}}[w]  =  - \ds{\sum_{\ell=1}^L} q_\ell w_\ell &  & \\
\text{subject to} &  I_{\ul{q}}(X_0;X_2) - I_{\ul{q}}(X_1;Y|X_0,X_2) & \leq & 0 \\
  &     -q_\ell  & \leq & 0 \\
 &     -1 + \ds{\sum_{\ell=1}^L} q_\ell  & =&  0 \\
& \forall x_0, \;  \sum_{\ell \in \mc{L}_{X_0}(x_0)} q_{\ell} -\rho(x_0)
 &=& 0 \\
& \forall (x_1,y), \
\frac{\sum_{\ell \in \mc{L}_{X_1,Y}(x_1,y)}
 q_{\ell} }{\sum_{\ell \in \mc{L}_{X_1}(x_1)} q_{\ell}}  - \Gamma(y|x_1) &=& 0
\end{array}
\end{equation}

where $q_\ell$ is the probability of a given quadruplet $(x_0,x_1,x_2,y)$,
$w_\ell$ is the value of the corresponding payoff, the vector $\ul{q} = (q_1, ..., q_L)$ represents the distribution $Q$, and the sets of indices
$\mc{L}_{X_0}(x_0)$, $\mc{L}_{X_1,Y}(x_1,y)$, $\mc{L}_{X_1}(x_1)$  merely translate the marginalization
conditions. By Lemma \ref{lemma:convexity-of-phi}, it follows that the above optimization problem is convex, which makes easy the determination of the information-constrained maximum of the expected payoff. A simple and useful upper bound for this maximum is $\mathbb{E}_{\rho} \max_{(x_1,x_2)} w(x_0,x_1,x_2)$. This bound will be referred to as the costless communication case in Sec. \ref{sec:application-PC-IC-numerical-results}. Indeed, this bound can be  attained in the ideal scenario where: given the knowledge of the coming state $x_0$, DM 1 computes an optimal solution for the action pair for the coming stage $(x_1^*,x_2^*) \in \arg \max_{(x_1,x_2)} w(x_0,x_1,x_2)$ and can inform DM 2 of $x_2^*$ without any cost. If the state is stationary for say $S$ stages and $\mc{X}_1=\mc{X}_2$, a simple strategy for DM 1 can be as follows: $x_1(1) = x_2^*$, $x_1(2)=x_1^*$, ..., $x_1(S) = x_1^*$. This allows DM 2 to choose an optimal action for $i\in\{2,...,S\}$. It can be shown that considering the $S-$stage block i.i.d. case amounts to multiplying the left term of (\ref{eq:information-constraint}) by $\frac{1}{S}$, which makes the info constraint arbitrarily mild as $S$ grows large.

\section{Application to power control over 
interference channels}
\label{sec:application-PC-IC-numerical-results}

\begin{figure}[htbp]
  \includegraphics[width=0.50\textwidth]{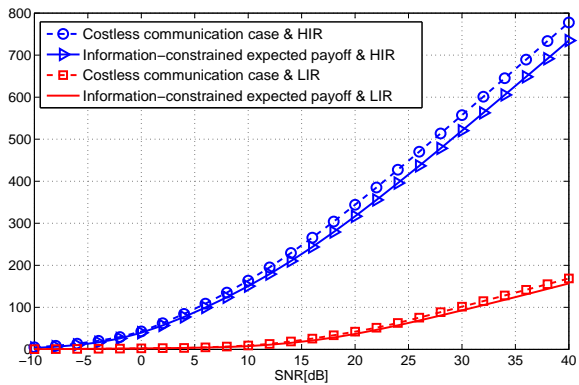}
\caption{Relative gain in terms of expected payoff (``OCPC/FPC - 1'' in [\%]) vs SNR[dB] obtained with CPC (with and without communication cost) when the reference power control policy is to transmit at full power (FPC).}
\label{fig1}
\end{figure}

\begin{figure}[htbp]
  \includegraphics[width=0.50\textwidth]{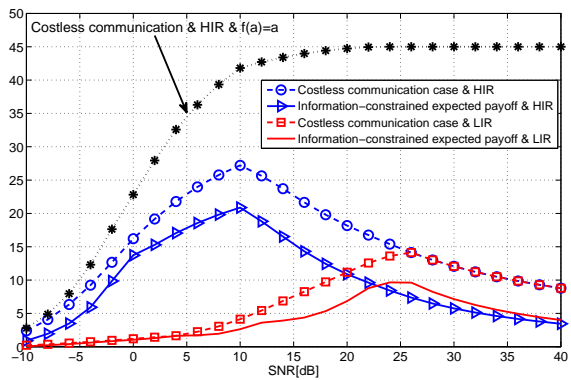}
\caption{The difference with Fig. \ref{fig1} is that the reference power control policy is the SPC policy. Additionally, the top curve is obtained with $f(a) =a$.}
\label{fig2}
\end{figure}

The main goal is to assess the performance of simple CPC policies and those of good policies, the performance of the latter is obtained by exploiting Theorem 1. A flat-fading interference channel (IC) with two Tx-Rx pairs is considered. Transmissions are assumed to be time-slotted and synchronized. For $j\in\{1,2\}$ and ``$k=-j$'' ($-j$ stands for the Tx other than $j$), the SINR at receiver $j$ at a given stage writes as $\mathrm{SINR}_j= \frac{g_{jj} x_j }{\sigma^2 + g_{kj} x_{k}}$
where $x_j \in \mathcal{X}_j^{\text{IC}} = \left\{0, P_{\max}\right\}$ is the power level chosen by Tx $j$, $g_{jk}$ represents the channel gain of link $jk$, and $\sigma^2$ the noise variance. We assume that: $g_{jk} \in \{g_{\min}, g_{\max}\}$ is i.i.d. and Bernouilli distributed $g_{jk} \sim \mc{B}(p_{jk})$ with $P(g_{jk} = g_{\min}) = p_{jk}$. We define $\text{SNR[dB]} = 10\log_{10}\frac{P_{\max}}{\sigma^2}$ and set $g_{\min} = 0.1$, $g_{\max}=1.9$, $\sigma^2=1$. The low and high interference regimes (LIR, HIR) are respectively defined by $(p_{11},p_{12},p_{21},p_{22}) = (0.5,0.9,0.9,0.5)$ and $(p_{11},p_{12},p_{21},p_{22}) = (0.5,0.1,0.1,0.5)$. The assumed payoff is $w^{\text{IC}}(x_0,x_1,x_2) =  \sum_{j=1}^2 f( \mathrm{SINR}_j(x_0,x_1,x_2))$ where $f(a) = \log(1+a)$ unless stated otherwise.  At last we assume that $Y\equiv X_1$. We consider four CPC policies~:\\
$\blacktriangleright$ the full power control (FPC) policy $x_j=P_{\max}$ for every stage. FPC requires no CSI at all;\\
$\blacktriangleright$ the semi-coordinated PC (SPC) policy $x_2=P_{\max}$, $x_1^{\dag} \in \arg \max_{x_1} w^{\text{IC}}(x_0, x_1, P_{\max})$. SPC requires the knowledge of the current state realization at Tx1;\\
$\blacktriangleright$ the optimal CPC policy (OCPC) whose performance are 
obtained, in particular, when the problem has the information structure of Theorem 1;\\
$\blacktriangleright$ the costless communication case (see Sec. \ref{sec:comments-main-result}) for which the maximum of $w^{\text{IC}}$ can be reached at any stage. Fig. \ref{fig1} and \ref{fig2} depict the relative gain in \% in terms of average payoff versus SNR[dB] which is obtained by FPC, SPC, OCPC, and costless case. Compared to FPC, gains are very significant whatever the interference regime and provided the SNR has realistic values. Compared to SPC, the gain is of course less impressive since SPC is precisely a coordinated PC scheme but, in the HIR and when the communication cost is negligible, gains as high as $25\%$ can be obtained with $f(a) = \log(1+ a)$ and $45\%$ with $f(a) = a$.

\section{Concluding remarks}
\label{sec:conclusion}

Although some assumptions made in this paper might be too restrictive
 in some application scenarios, it is essential to understand that the used methodology to derive the optimal performance is general. It can be applied to analyze the performance of coded power allocation, coded interference alignment, etc, with other information structures and by considering $N\geq 2$ individual payoffs instead of a common one (e.g., in a game-theoretic setting \cite{Gossner-2006}). The methodology to assess the performance of good coded policies consists in deriving the right information constraint(s) by building the proof on Shannon theory for the problem of multi-source coding with distortion over multi-user channels wide side information and then to use this constraint to find an information-constrained maximum of the payoff (common payoff case) or the set of Nash equilibrium points which are compatible with the constraint (non-cooperative game case). Note that assuming i.i.d. from stage to stage the state(s) leads in fact to the worst-case scenario for the information constraint. On the other hand, the costless communication case provides an upper bound for the expected payoff. As a key observation, the communication structure of a multi-person decision-making problem is a multiuser channel, which makes multi-terminal Shannon theory not only relevant for pure communication problems but also for any multi-person decision-making problem. This observation opens new challenges for Shannon-theorists since decision-making problems define new channels for instance. It can also be observed that the need to design payoff-oriented communications urges a rethinking of the problem of coding.


%
%
%
%
%
%
%
%
%
%
%
%
%

\section*{Acknowledgment}
The authors would like to thank Prof. Olivier Gossner for interesting feedbacks
on his work.

\bibliographystyle{IEEEtran}
\bibliography{biblio-v2}
%

\end{document}